\documentclass[a4paper]{jpconf}
\usepackage{graphicx}
\usepackage{amssymb}
\usepackage{xcolor}
\usepackage{cite}

\newcommand{\NREC}{N_{\Lambda \,\mbox{\tiny{REC}}}}
\newcommand{\NQGP}{N_{\Lambda \,\mbox{\tiny{QGP}}}}
\newcommand{\NbarREC}{N_{\overline{\Lambda} \,\mbox{\tiny{REC}}}}
\newcommand{\NbarQGP}{N_{\overline{\Lambda} \,\mbox{\tiny{QGP}}}}
\newcommand{\NupREC}{N^\uparrow_{\Lambda \,\mbox{\tiny{REC}}}}
\newcommand{\NdownREC}{N^\downarrow_{\Lambda \,\mbox{\tiny{REC}}}}
\newcommand{\NupQGP}{N^\uparrow_{\Lambda \,\mbox{\tiny{QGP}}}}
\newcommand{\NdownQGP}{N^\downarrow_{\Lambda \,\mbox{\tiny{QGP}}}}
\newcommand{\NbarupREC}{N^\uparrow_{\overline{\Lambda} \,\mbox{\tiny{REC}}}}
\newcommand{\NbardownREC}{N^\downarrow_{\overline{\Lambda} \,\mbox{\tiny{REC}}}}
\newcommand{\NbarupQGP}{N^\uparrow_{\overline{\Lambda} \, \mbox{\tiny{QGP}}}}
\newcommand{\NbardownQGP}{N^\downarrow_{\overline{\Lambda} \,\mbox{\tiny{QGP}}}}

\newcommand{\NpQGP}{N_{p \,\mbox{\tiny{QGP}}}}

\newcommand{\PLambda}{\mathcal{P}^{\Lambda}}
\newcommand{\PLambdabar}{\mathcal{P}^{\overline{\Lambda}}}
\newcommand{\PLambdaREC}{\mathcal{P}^{\Lambda}_{\mbox{\tiny{REC}}}}
\newcommand{\PLambdabarREC}{\mathcal{P}^{\overline{\Lambda}}_{\mbox{\tiny{REC}}}}

\begin{document}
\title{Two-component source to explain $\Lambda$ and $\bar{\Lambda}$ global polarization in non-central heavy-ion collisions}

\author{A Ayala$^{1,2}$, E Cuautle$^1$, I Dom\'{\i}nguez$^3$, I Maldonado$^3$, J Salinas$^1$ and M E Tejeda-Yeomans$^{4,5}$} 

\address{$^1$Instituto de Ciencias Nucleares, Universidad Nacional Aut\'onoma de M\'exico, Apartado
  Postal 70-543, CdMx 04510,  Mexico}
\address{$^2$Centre for Theoretical and Mathematical Physics, and Department of Physics, University of Cape Town, Rondebosch 7700, South Africa}
\address{$^3$Facultad de Ciencias F\'isico-Matem\'aticas, Universidad Aut\'onoma de Sinaloa, Avenida de las Am\'ericas y Boulevard Universitarios, Ciudad Universitaria, C.P. 80000, Culiac\'an, Sinaloa, Mexico}
\address{$^4$Facultad de Ciencias - CUICBAS, Universidad de Colima, Bernal D\'iaz del Castillo No. 340, Col. Villas San Sebasti\'an, 28045 Colima, Mexico}
\address{$^5$Departamento de Física, Universidad de Sonora, Boulevard Luis Encinas J. y Rosales, Colonia Centro, Hermosillo, Sonora 83000, Mexico}

\ead{ivonne.alicia.maldonado@gmail.com}

\begin{abstract}
	The STAR Beam Energy Scan program has found a difference in the global polarization of $\Lambda$s and $\bar{\Lambda}$s produced in Au + Au collisions. This difference is larger for lower center of mass collision energies.
	In this work we show that a two-component source, consisting of a high density core and a lower density corona, can describe quantitatively the $\Lambda$ and $\bar{\Lambda}$ polarizations as a function of collision energy.
\end{abstract}

\section{Introduction}

The polarization of hadrons in high energy collisions has been a topic of much interest mainly because this is related to the possibility to understand the role played by spin degrees of freedom during hadronization. In heavy-ion reactions, hadron polarization properties can provide useful information to understand hadron  production mechanisms. The polarization of particles along a preferred direction sheds light on how these properties are linked to those  of the medium produced in the reaction, which in turn are associated with the initial conditions and the dynamics of the Quark Gluon Plasma (QGP). In these reactions, the possibility to create a vortical fluid can be quantified in terms of the thermal vorticity \cite{Jacob:1988dc,Barros,Ladygin,Becattini1,Xie,Karpenko,Xie2,Liao,Liao2,Li, Karpenko2,Xia,Suvarieva}.

The $\Lambda$ polarization was discovered in the mid-70's\cite{Bunce} and it has been studied in different collision systems such as p+p, p+A, A+A, DIS and e$^+$ + e$^-$ over a wide energy range. 
In this context, the $\Lambda$ hyperon is one of the most studied particles given that it is easy to produce and to identify by reconstructing the topology of its decay into two particles of opposite charge. However, the mechanism whereby polarization occurs has not yet been understood, although several models have been developed to try to explain it  \cite{Groom,Yen,Andersson,DeGrand,Panagiotou,Soffer,Gustafson,Ellis,Ellis2,Jaffe,Kotzinian,Florian,Florian2,Boros,Anselmino,Anselmino2,Ma,Alikhanov,Yang,Sun,Han}. Furthermore, interest in this topic has been recently renewed since the global polarization can be used to measure the thermal vorticity and perhaps also the electromagnetic field generated within the plasma stage in  non-central heavy-ion collisions~\cite{Becattini2008,Becattini2017}.

Recently the magnitude of the global polarization has been measured by ALICE\cite{Acharya:2019ryw} and the STAR Beam Energy Scan (BES)\cite{STAR-Nature,STAR2}. The results of the latter  have obtained the global polarization of $\Lambda$s and $\bar{\Lambda}$s as a function of the  collision energy. These polarization are at the level of a few percent and decrease as the collision energy increases from $\sqrt{s_{NN}} =$ 7.7 to 200 GeV. They also exhibit a clear difference between $\Lambda$ and $\bar{\Lambda}$ polarizations, the latter being larger for lower collision energies. 

In this work we show that the $\Lambda$ and $\bar{\Lambda}$ polarization in semi-central heavy-ion collisions can be understood when accounting for the abundances of $\Lambda$s and $\bar{\Lambda}$s  coming from a high density core and a less dense corona, using the results of the intrinsic $s$-quark/antiquark polarization as a seed for the $\Lambda$/$\bar{\Lambda}$ polarization~\cite{Ayala:2020soy}.

\section{$\Lambda$ and $\bar{\Lambda}$ from two-component model}

\begin{figure}[h]
\includegraphics[width=14pc]{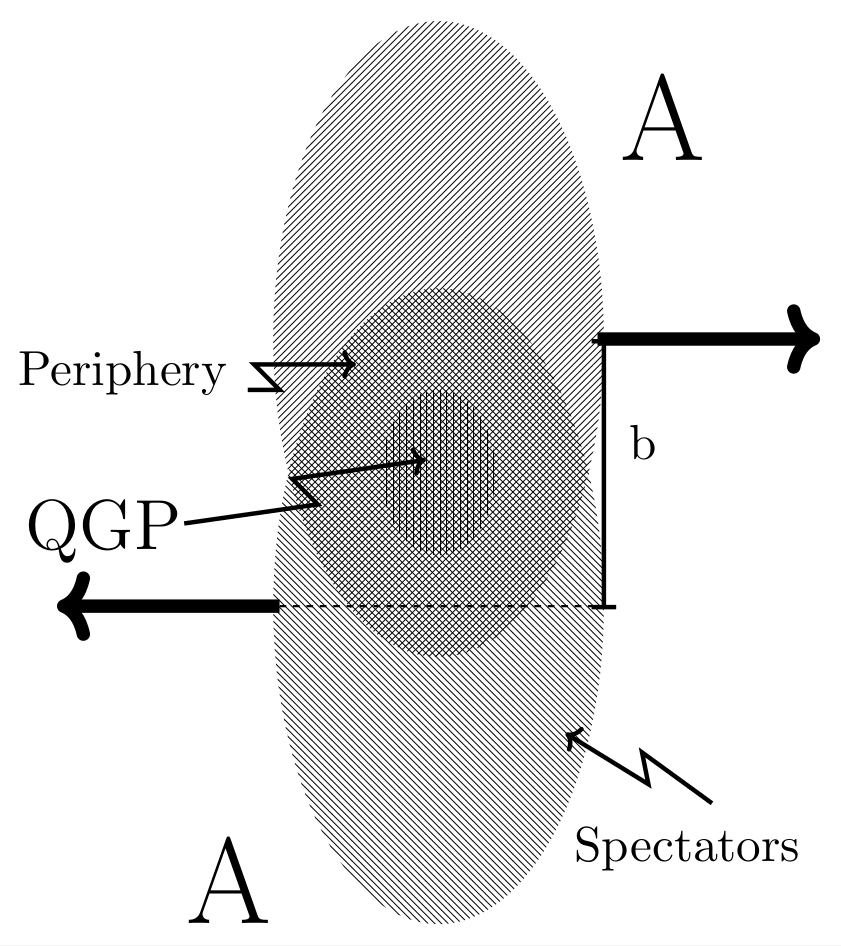}\hspace{1.5pc}%
	\begin{minipage}[b]{22pc}\caption{\label{figura1} Representation of a  non-central heavy-ion collision of a symmetric system with impact parameter $b$(Taken from Ref. \cite{ayalacuautle}). The $\Lambda$s and $\bar{\Lambda}$s coming from the core or central region are produced by QGP processes,whereas, those from the periphery, or corona, are produced by n + n kind of reactions.}
\end{minipage}
\end{figure}

In non-central heavy ion collisions, $\Lambda$s and $\bar \Lambda$s are produced in different density regions within the interaction zone~\cite{ayalacuautle}, as is shown in the Fig.~\ref{figura1}. Therefore, their properties can differ depending on whether these particles come from the core, the central region with higher density, or from the corona, the peripheral region with lower density. 

\noindent In this scenario the number of $\Lambda$s produced in the collision $N_\Lambda$ can be rewritten in terms of the number of $\Lambda$s coming from the corona $\NQGP$ plus the number of $\Lambda$s originating from the core $\NREC$, such that
\begin{equation}
	N_\Lambda=\NQGP + \NREC
\end{equation}
\noindent The subscripts ``QGP'' and ``REC'' refer to the kind of processes responsible for the production of these hyperons, that is, QGP and recombination induced processes, respectively. The latter are similar to the polarization producing processes in p+p reactions.

\section{Polarization from a two-component model}

\noindent Considering that the polarization asymmetry of a given baryon species produced in high energy reactions, defined by
\begin{equation}
	\PLambda = \frac{N^\uparrow - N^\downarrow}{N^\uparrow + N^\downarrow}
\end{equation}
\noindent namely, the ratio of the difference between the number of baryons with their spin pointing along and opposite to a given direction, to their sum, can be rewritten in terms of the number of $\Lambda$s or $\bar{\Lambda}$s produced in the different density regions, then the $\Lambda $ and $\bar{\Lambda}$ polarization is given by 

\begin{eqnarray}
	\PLambda&=&\frac{\
(\NupQGP+\NupREC)-(\NdownQGP+\NdownREC)}{\ 
(\NupQGP+\NupREC)+(\NdownQGP+\NdownREC)},\nonumber \\
	\PLambdabar&=&\frac{\ 
(\NbarupQGP+\NbarupREC)-(\NbardownQGP+\NbardownREC)}{\ 
(\NbarupQGP+\NbarupREC)+(\NbardownQGP+\NbardownREC)}.
\label{polLambda}
\end{eqnarray}

\noindent After some algebra, we can express the previous equations as
\begin{eqnarray}
	\PLambda &=&
\frac{\left( \PLambdaREC +
\frac{\NupQGP - \NdownQGP}{\NREC}\right)}{ \left( 1 + \frac{\NQGP}{\NREC}\right)}, \nonumber\\
	\PLambdabar&=&\frac{\left( \PLambdabarREC + \frac{\NbarupQGP - \NbardownQGP}{\NREC}\right)}{ \left( 1 + \frac{\NbarQGP}{\NbarREC}\right)},
	\label{polLambda2}
\end{eqnarray}
where $\PLambdaREC$ and $\PLambdabarREC$ refers to the polarization along the global angular momentum produced in the corona
\begin{eqnarray}
	\PLambdaREC &=&\frac{\NupREC - \NdownREC}{\NupREC + \NdownREC},\\
	\PLambdabarREC&=&\frac{\NbarupREC - \NbardownREC}{\NbarupREC + \NbardownREC}.
\label{PREC}
\end{eqnarray}

\noindent We can make some working estimations as follows: 
\begin{itemize}
	\item $\PLambdaREC= \PLambdabarREC = 0$ because, although the colliding nucleons  in this zone partake of the vortical motion, reactions in cold nuclear matter are less efficient to align the spin in the direction of the angular momentum than in the QGP.

	\item The intrinsic global $\Lambda$ and $\bar \Lambda$ polarizations $z$ and $\bar z$, defined as
\begin{eqnarray}
    z&=& \frac{(\NupQGP- \NdownQGP)}{\NQGP} \nonumber\\
   \bar{z}&=& \frac{(\NbarupQGP - \NbardownQGP)}{ \NbarQGP}\simeq \frac{(\NbarupQGP - \NbardownQGP)}{\NQGP},
\label{aprox2}
\end{eqnarray}
are non zero and come mainly from the core, since the reactions in this zone are more efficient to align the particle spin to global angular momentum, albeit they are small. Additionally, we assume that $\NbarQGP \simeq \NQGP$ in the QGP.

\item We expect that the number of $\Lambda$s produced in the corona is larger than the number of $\bar{\Lambda}$s, since they come from cold nuclear matter collisions and these processes are related to p + p reactions in which it is more difficult to produce three antiquarks than only one $s$-quark. Then we can write 
	\begin{equation}
		\NbarREC \equiv w \NREC
	\end{equation}
with $w < 1$. Simulations of p + p collisions using the Ultrarelativistic Quantum Molecular Dynamics (UrQMD) generator, show that the value of $w$ is smaller than 0.4 for the energies considered as can be seen in Fig.~\ref{dobleu}. 
\begin{figure}[h]
\includegraphics[width=18pc]{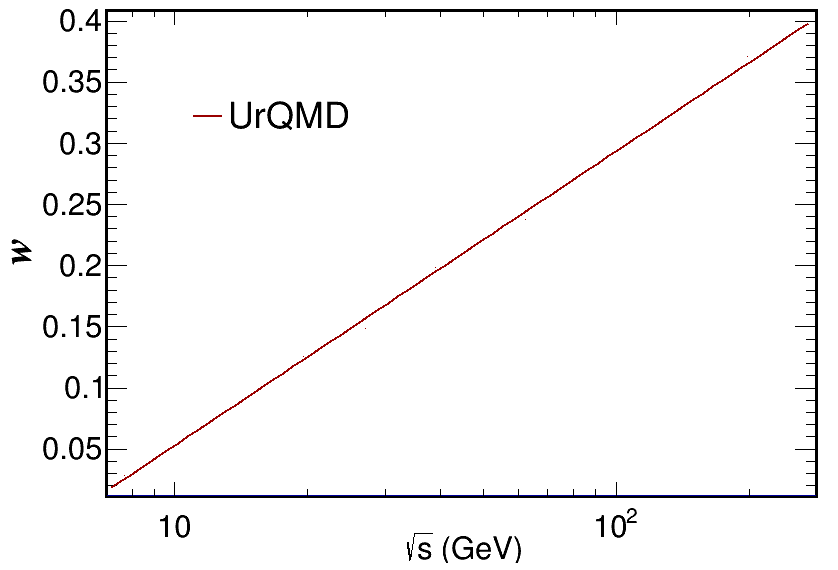}\hspace{1.5pc}%
	\begin{minipage}[b]{18pc}\caption{\label{dobleu} $w = \NbarREC/\NREC$ from p + p collisions simulated with UrQMD. $w$ increases with energy. For energy ranges below 200 GeV, the maximum $w$-value is smaller than 0.4}
\end{minipage}
\end{figure}
\end{itemize}

\begin{figure}[h]
	\includegraphics[width=16pc]{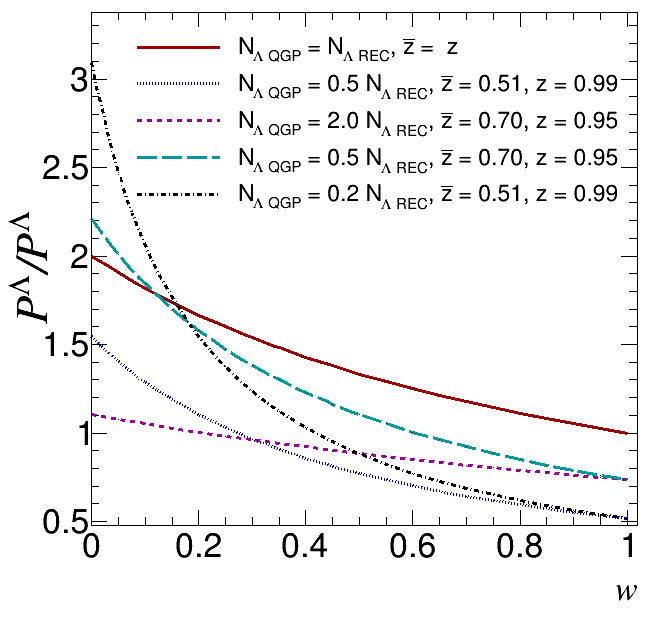}\hspace{1.5pc}%
	\begin{minipage}[b]{20pc}\caption{\label{figure6} $\PLambdabar/\PLambda$ as a function of $w$. We observe that this ratio decreases as $w$ increases. However, there is a range of $w$-values for which $\PLambdabar/\PLambda > 1$ whose width depends on the relation between $\NQGP$, $\NREC$, $z$ and $\bar{z}$.}
\end{minipage}
\end{figure}

\noindent With the previous assumptions we can rewrite Eqs.~\ref{polLambda2} as

\begin{eqnarray}
	\PLambda &=& \frac{z\frac{\NQGP}{\NREC}}{\left(1 + \frac{\NQGP}{\NREC} \right)}, \\ 
	\PLambdabar &=&\frac{\left( \frac{\bar{z}}{w}\right) \frac{\NQGP}{\NREC}}{\left(1 + \left(\frac{1}{w} \right)\frac{\NQGP}{\NREC} \right)}
	\label{polLambda3}
\end{eqnarray}
which depend on $z$ and $\bar z$ the intrinsic $\Lambda$ and
$\bar{\Lambda}$ polarization, $w$ and the ratio
$\frac{\NQGP}{\NREC}$. We expect $\bar{z}$ to be smaller than $z$ Nonetheless, the factor $1/w > 1$ amplifies the
global $\bar{\Lambda}$ polarization which leads to $\PLambdabar >
\PLambda$. This is shown in Fig. \ref{figure6} in which we
plot $\PLambdabar/\PLambda$ as a function of $w$, for different values of the
intrinsic polarization and of the ratio of $\Lambda$ and $\bar{\Lambda}$ polarizations. 
We can see that in the extreme situation where $\bar{z} = z$ and
$\NQGP/\NREC=1$, $\PLambdabar/\PLambda$ is always larger than 1 for $0<w<1$. Considering a more realistic scenario, we take the intrinsic polarization to satisfy $\bar{z}<z$
with $\NQGP/\NREC$ smaller than 1 and find that there is still a range of $w$ values for which $\PLambdabar/\PLambda$ is larger than 1. This region shrinks when $\NQGP/\NREC>1$.

To get the ratio $\NQGP/\NREC$ we analyze the production of $\Lambda$s in the QGP and REC regions. The number of $\Lambda$s in the QGP region varies as 
the square of the number of participants $\NpQGP$ in the collision  at a given impact parameter $b$

\begin{equation}
	\left< s \right > = \NQGP = c \NpQGP^2
\end{equation}

\noindent with $0.001 \leq c \leq 0.005$ \cite{Letessier}, for the case when assuming that all the $s$-quarks hadronize into $\Lambda$s. Nevertheless, since we know that also other strange hadrons are produced, $c$ becomes smaller. 

The number of participants is calculated with 

\begin{equation}
	\NpQGP =\int n_p(\vec{s}, \vec{b})\theta[n_p( \vec{s}, \vec{b}) - n_c] d^2s
\end{equation}
in terms of the density of participants per unit transverse area given by
\begin{equation}
	n_p(\vec{s},\vec{b}) = T_{A}(\vec{s})[1-e^{-\sigma_{NN}T_{B}(\vec{s}-\vec{b})}] + T_{B}(\vec{s}-\vec{b})[1 - e^{-\sigma_{NN}T_{A}(\vec{s})}]
\end{equation}
with $\vec{b}$ the vector directed along the impact parameter on the nuclei overlap area, $\sigma_{NN}$ is the collision energy dependent nucleon-nucleon cross-section~\cite{PDG, ALICE:sigmaNN} and $T_{A}(\vec{s}) = \int_{-\infty}^{\infty}\rho_{A}(\vec{s},z)dz$ is the thickness function, the nucleon density per unit area in the transverse plane to the collision axis. We take as the nuclear density 
\begin{equation}
	\rho_A(\vec r) = \frac{\rho_0}{1+e^{\frac{r-R_{A}}{a}}}
\end{equation}
a Woods-Saxon profile with a skin depth $a=0.41$ and a radius for a nucleus with mass number A of $R_A = 1.1A^{1/3}$ fm.

\begin{figure}[h]
	\includegraphics[width=16pc]{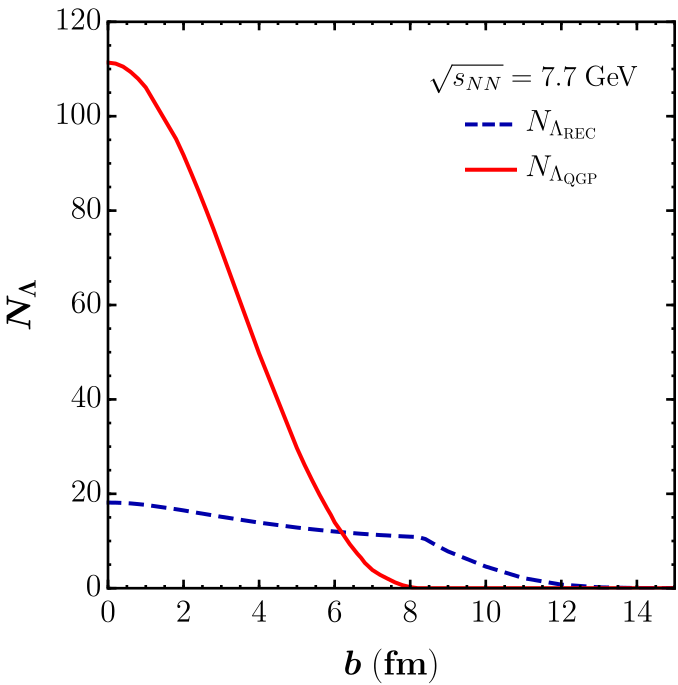}\hspace{1.5pc}%
	\begin{minipage}[b]{20pc}\caption{\label{figura7} Number of $\Lambda$s produced in the core and in the corona as a function of impact parameter. The number of $\Lambda$s produced in the core $\NQGP$ is smaller than the number of $\Lambda$s produced in the corona for an impact parameter larger than a certain value. In the case of $\sqrt{s_{NN}} = 7.7$ GeV, this occurs for $b \gtrsim 6$ fm. }
\end{minipage}
\end{figure}

On the other hand, the number of $\Lambda$s produced in the corona can be written as
\begin{equation}
	\NREC = \sigma_{NN}^{\Lambda} \int d^2s T_{B}(\vec{b}-\vec{s})T_{A}(\vec{s}) \theta[n_c -n_p(\vec{s},\vec{b})]
\end{equation}

\noindent where the nucleon-nucleon cross-section for $\Lambda$ production is parameterized using data as $\sigma_{\Lambda}^{NN}= 1.37 \ln(\sqrt{s} -0.94)$. As in the case of the QGP region, the number of $\Lambda$s depends on the minimal critical density  $n_c$ required to produce a QGP. The accepted value is $3.3$ fm$^{-2}$ given by the $J/\Psi$ suppression in the threshold model~\cite{Blaizot:1996nq}. Figure ~\ref{figura7} shows $\NREC$ and $\NQGP$ as a function of impact parameter for a collision energy of $\sqrt{s_{NN}} = 7.7$ GeV. Notice that in this case, the ratio $\NQGP/\NREC$ becomes smaller than 1 for $b \gtrsim 6$ fm.      

\noindent We consider that the intrinsic global polarization of $\Lambda$ and $\bar{\Lambda}$ comes from the corresponding polarization of strange quarks ($z$) and antiquarks ($\bar{z}$)~\cite{WWND2020:Ayala} given by
\begin{eqnarray}
	z = 1 - e^{-\frac{ t}{\tau}}, \  	\bar{z} = 1 - e^{-\frac{ t}{\bar{\tau}}}
	\label{intr}
\end{eqnarray}
 as a function of the formation time ($t$) of $\Lambda$ ($\bar \Lambda$) and the relaxation time $\tau$ ($\bar{\tau}$) required for the alignment between the spin of the quark (antiquark) and the thermal vorticity. We resort to the results of Ref. \cite{relaxtime} in which the relaxation time, for the alignment between the spin of  quark $s$ or $\bar{s}$ with the thermal vorticity is computed as a function of the collision energy calculated at the freeze-out, when the maximum baryon density is reached, in terms of the temperature $T$ and the baryon chemical potential $\mu_B$ as a function of the energy parameterized as
\begin{eqnarray}
	T(\mu_B) &=& 166 - 139 \mu^2_B - 53 \mu^4_B \nonumber \\
	\mu_B &=& \frac{1308}{1000 + 0.273\sqrt{s}}
\end{eqnarray}
where $\mu_B$ and $T$ are expressed in MeV\cite{Randrup:2006nr}.

\begin{figure}[h]
\includegraphics[width=16pc]{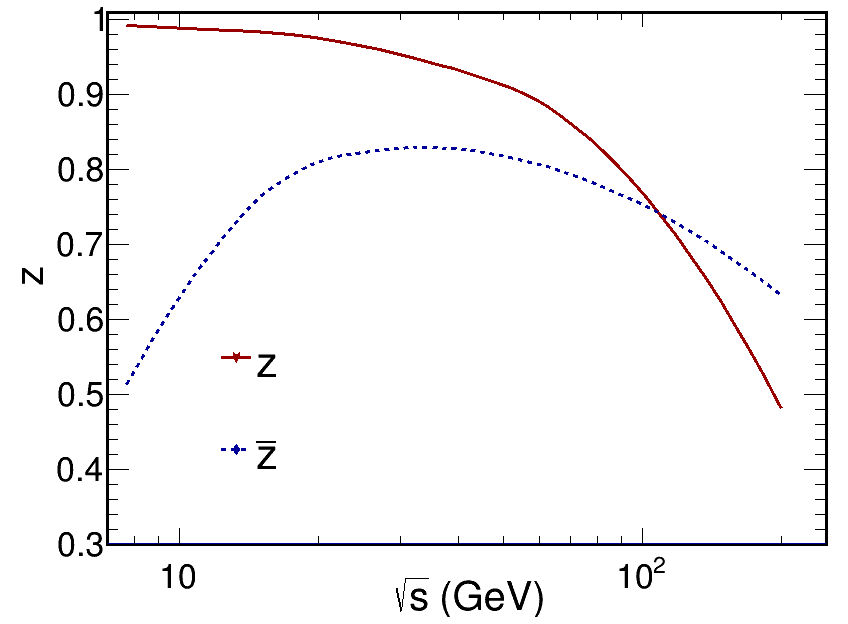}\hspace{1.5pc}%
	\begin{minipage}[b]{20pc}\caption{\label{zparameter}
		Intrinsic global polarization $z$ for $\Lambda$ and $\bar{z}$ for $\bar{\Lambda}$ as functions of energy $\sqrt{s_{NN}}$ for semicentral collisions at an impact parameter $b = 8$ fm and $t = 5$ fm. Notice that $\bar{z} < z$ for $\sqrt{s_{NN}} < 200$ GeV, energy for which this behavior is inverted.}
\end{minipage}
\end{figure}

Figure \ref{zparameter} shows the intrinsic global polarization $z$ for $\Lambda$ and $\bar{z}$ for $\bar{\Lambda}$ calculated with Eq.~(\ref{intr}) for semi-central collisions at an impact parameter $b = 8$ fm and a formation time $t = 5$ fm, as a function of energy. We observe that $\bar{z} < z$ for low collision energies, behavior that is inverted for energy $\sqrt{s_{NN}} > 200$ GeV. For smaller formation times the intrinsic polarization $z$ and $\bar{z}$ are smaller but the trend is similar. 

\begin{figure}[h]
\begin{minipage}{36pc}
\includegraphics[width=17pc]{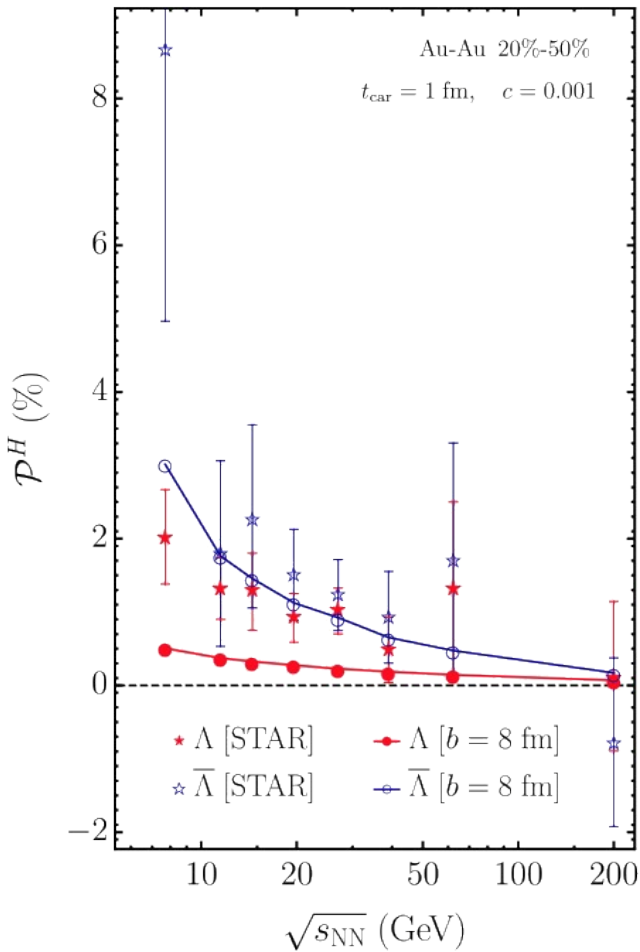}
\includegraphics[width=17pc]{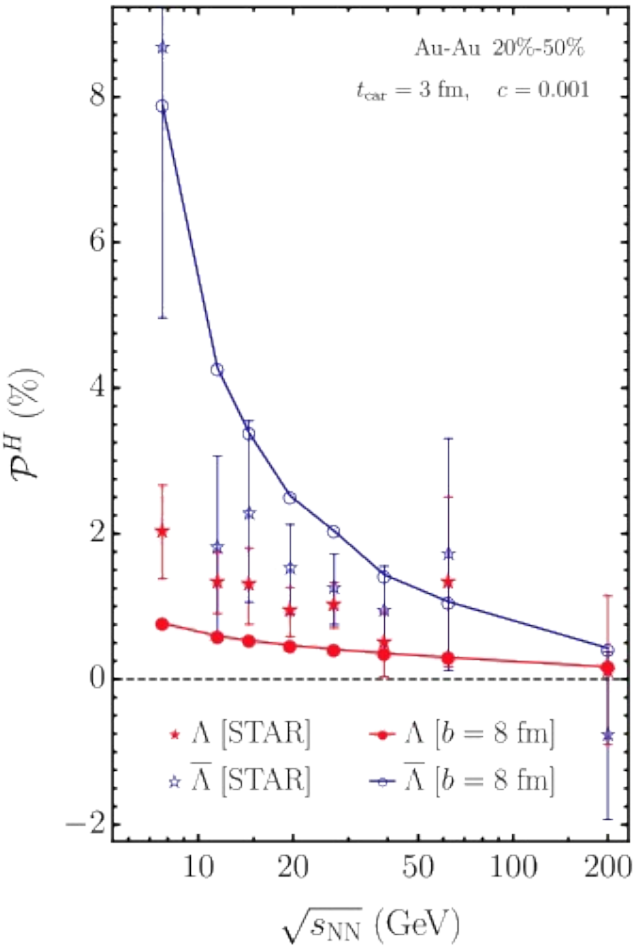}
	\caption{\label{label-1}Global $\Lambda$ and $\bar{\Lambda}$ polarizations  compared with data from BES~\cite{STAR-Nature} for two different formation times $t = 1$, and  $3$ fm.}
\end{minipage} 
\end{figure}

Figure \ref{label-1} shows the global polarization calculated for two different formation times for $\Lambda$ and $\bar{\Lambda}$ within the QGP, $t = {1,3}$ fm and for $b= 8$ fm. The calculation is compared with data from the STAR-BES. Notice that the data is well described over the entire collision energy range.

\section{Summary and Conclusions}

The $\Lambda$ and $\bar{\Lambda}$ global polarization measured by the STAR-BES shows that the polarization of $\bar{\Lambda}$ is larger than $\Lambda$ at low energies and the difference decreases as the energy increases. In this work we have shown that this can be understood in a two component model where $\Lambda$s come from a high density core, namely, the region where the QGP is created, and a less dense corona where the production mechanism comes from p+p like processes. 
We have shown that when the ratio $N_{\bar{\Lambda}\,\mbox{\tiny{REC}}}/N_{\Lambda \,\mbox{\tiny{REC}}}$ is smaller than one, an amplification for the global $\bar{\Lambda}$ polarization over that of $\Lambda$ is obtained, in spite of the intrinsic $\Lambda$ being larger than the intrinsic $\bar{\Lambda}$ polarization. With reasonable assumptions, this two-component model provides a 
qualitative and quantitative description of the STAR-BES data. A more detailed analysis has been performed and reported in \cite{Ayala:2020soy}.

\section*{Acknowledgments}

A.A. thanks F. Wang for helpful comments. Support for this work has been received in part by UNAM-DGAPA-PAPIIT grant number IG100219 and by Consejo Nacional de Ciencia y Tecnolog\'{\i}a grant numbers A1-S-7655 and  A1-S-16215. I.M. acknowledges support from a postdoctoral fellowship granted by Consejo Nacional de Ciencia y Tecnolog\'{\i}a.

\end{document}